\documentclass{aa} 
\usepackage{natbib}
\usepackage{graphicx}
\usepackage{amssymb}
\usepackage{epsfig}
\usepackage{graphicx}
\usepackage{txfonts}

\def\eg{{\it e.g.\ }}

\def\ie{{\it i.e.\ }}
\def\vs{{\it versus\ }}
\begin{document}
\title{The VIRMOS deep imaging survey IV: \\Near-infrared observations \thanks{Based on observations collected at the European Southern Observatory, La Silla, Chile. The data discussed in this paper will be made available to the astronomical community at the following link: http://cencosw.oamp.fr/.}}  

\author{
A. Iovino \inst{1} 
\and H.J. McCracken \inst{2,3}
\and B. Garilli \inst{4}
\and S. Foucaud \inst{4}
\and O. Le F\`evre \inst{5}
\and D. Maccagni \inst{4}
\and P. Saracco \inst{1}
\and S. Bardelli  \inst{6}
\and G. Busarello \inst{7} 
\and M. Scodeggio \inst{4}
\and A. Zanichelli \inst{8}
\and L. Paioro \inst{4} 
\and D. Bottini \inst{4}
\and V. Le Brun \inst{5}
\and J.P. Picat \inst{9}
\and R. Scaramella \inst{8}
\and L. Tresse \inst{5}
\and G. Vettolani \inst{8}
\and C. Adami \inst{5}
\and M. Arnaboldi \inst{7}
\and S. Arnouts \inst{5}
\and M. Bolzonella  \inst{10}
\and A. Cappi    \inst{6}
\and S. Charlot \inst{2,11}
\and P. Ciliegi    \inst{6}
\and T. Contini \inst{9}
\and P. Franzetti \inst{4}
\and I. Gavignaud \inst{9,12}
\and L. Guzzo \inst{1}
\and O. Ilbert \inst{10}
\and B. Marano  \inst{10}
\and C. Marinoni \inst{1}
\and A. Mazure \inst{5}
\and B. Meneux \inst{5}
\and R. Merighi  \inst{6}
\and S. Paltani \inst{5}
\and R. Pell\`o \inst{9}
\and A. Pollo \inst{1}
\and L. Pozzetti    \inst{6}
\and M. Radovich \inst{7}
\and G. Zamorani \inst{6}
\and E. Zucca    \inst{6}
\and E. Bertin \inst{2,3}
\and M. Bondi \inst{8}
\and A. Bongiorno \inst{10}
\and O. Cucciati \inst{1,13}
\and L. Gregorini \inst{8}
\and G. Mathez \inst{9}
\and Y. Mellier \inst{2,3}
\and P. Merluzzi \inst{7}
\and V. Ripepi \inst{7}
\and D. Rizzo \inst{9}
}

\institute{
INAF-Osservatorio Astronomico di Brera - Via Brera 28, Milan,
Italy
\and
Institut d'Astrophysique de Paris, UMR 7095, 98 bis Bvd Arago, 75014
Paris, France
\and
Observatoire de Paris, LERMA, 61 Avenue de l'Observatoire, 75014 Paris,
France
\and
IASF-INAF - via Bassini 15, I-20133, Milano, Italy
\and
Laboratoire d'Astropysique de Marseile, UMR 6110 CNRS-Universit\'e de
Provence,  BP8, 13376 Marseille Cedex 12, France
\and
INAF-Osservatorio Astronomico di Bologna - Via Ranzani,1, I-40127, Bologna, Italy
\and
INAF-Osservatorio Astronomico di Capodimonte - Via Moiariello 16, I-80131, Napoli,
Italy
\and
IRA-INAF - Via Gobetti,101, I-40129, Bologna, Italy
\and
Laboratoire d'Astrophysique de l'Observatoire Midi-Pyr\'en\'ees (UMR
5572) - 14, avenue E. Belin, F31400 Toulouse, France
\and
Universit\`a di Bologna, Dipartimento di Astronomia - Via Ranzani,1,
I-40127, Bologna, Italy
\and
Max Planck Institut fur Astrophysik, 85741, Garching, Germany
\and
European Southern Observatory, Karl-Schwarzschild-Strasse 2, D-85748
Garching bei Munchen, Germany
\and
Universit\'a di Milano-Bicocca, Dipartimento di Fisica, Piazza delle Scienze, 3, 
I-20126 Milano, Italy 
}

   \offprints{A. Iovino, iovino@brera.mi.astro.it}

\authorrunning{Iovino et al.}
\titlerunning{The VIRMOS deep imaging survey IV: Near-infrared observations}


\abstract{In this paper we present a new deep, wide-field
  near-infrared imaging survey. Our $J-$ and $K-$band observations in
  four separate fields (0226-04, 2217+00, 1003+02, 1400+05) complement
  optical $BVRI$, ultraviolet and spectroscopic observations
  undertaken as part of the VIMOS-VLT deep survey (VVDS). In total,
  our survey spans $\sim 400~\rm{arcmin}^2$. Our catalogues are
  reliable in all fields to at least $K\sim20.75$ and $J\sim21.50$
  (defined as the magnitude where object contamination is less than
  10\% and completeness greater than 90\%).
  
  Taken together these four fields represents a unique combination of
  depth, wavelength coverage and area. Most importantly, our survey
  regions span a broad range of right ascension and declination which
  allow us to make a robust estimate of the effects of cosmic
  variance. We describe the complete data reduction process from raw
  observations to the construction of source lists and outline a
  comprehensive series of tests carried out to characterise the
  reliability of the final catalogues. From simulations we determine
  the completeness function of each final stacked image, and estimate
  the fraction of spurious sources in each magnitude bin. We compare
  the statistical properties of our catalogues with literature
  compilations. We find that our $J-$ and $K-$selected galaxy counts
  are in good agreement with previously published works, as are our
  $(J-K)$ versus $K$ colour-magnitude diagrams. Stellar number counts
  extracted from our fields are consistent with a synthetic model of
  our galaxy. Using the location of the stellar locus in
  colour-magnitude space and the measured field-to-field variation in
  galaxy number counts we demonstrate that the \textit{absolute}
  accuracy of our photometric calibration is at the $5\%$ level or
  better. Finally, an investigation of the angular clustering of $K-$
  selected extended sources in our survey displays the expected
  scaling behaviour with limiting magnitude, with amplitudes in each
  magnitude bin in broad agreement with literature values. 
 
  In summary, these catalogues will be an excellent tool to
  investigate the properties of near-infrared selected galaxies, and
  such investigations will be the subject of several articles
  currently in preparation.

 }

   \maketitle
%
%
%
\newpage 

\section{Introduction}\label{intro}

Galaxy surveys selected in near-infrared wavelengths ($1.1-1.5\mu$)
provide some well-established advantages with respect their
optically-selected counterparts. A census conducted at these longer
wavelengths can provide flux measurements in an object's rest-frame
optical bandpass at intermediate redshifts, which can in turn be
easier to relate to physically interesting quantities such a galaxy's
total mass in stars. Any credible model of galaxy formation must
predict how this stellar mass function evolves with redshift. In
addition, the predicted number densities and spatial distribution of
objects lying at the reddest outer reaches of the optical-infrared
colour-magnitude diagram ( ``the extremely red objects'' or EROs) also
depend very sensitively on one's assumed model of galaxy
formation. Understanding this red outlier population and how it
relates to UV-selected star-forming galaxies at higher redshifts and
massive ellipticals at the present day has become one of the most
important questions in observational cosmology. Near-infrared data is
also crucial to compute accurate photometric redshifts in the $1<z<2$
redshift range, where measuring spectroscopic redshifts with a
red-optimised spectrograph can be challenging.  Until recently,
however, the small format of near-infrared detectors (and the much
higher ground--based brightness of the sky at longer wavelengths) has
made surveys of the near-infrared selected Universe a very
time-consuming undertaking.  Only in the last few years, with the
advent of larger-format detectors, has it become practical to survey
deeply larger areas of the sky in the near IR to cosmologically
significant redshifts.

In this paper we describe a new deep near-infrared survey. The
observations presented here cover a total area of $\sim
400$~arcmin$^2$ over four separate fields in both $J$ and $K$ bands.
Each of the four fields reaches a completeness limit (defined as the
magnitude at which $90\%$ of simulated point sources are recovered
from the images) of at least $22.0$ magnitudes in $J$ and 20.75 in
$K$.  This represents an intermediate regime between, for example,
very deep surveys like FIRES \citep{Labbe_et_al.2003.AJ} which covers
a few square arcminutes to depth of $K\sim24$, and shallower surveys
\citep{Daddi_et_al.2000b.A&A, Drory_et_al.2001.ApJ} which reach to
around $K\sim19$ over several hundred square arcminutes. The current
survey has been undertaken in the context of the VIMOS-VLT deep survey
\citep{Lefevre_et_al.2004.astro.ph}, and the near-infrared data
presented here complements optical and ultraviolet imagery
\citep{McCracken_et_al.2003.A&A, Radovich_et_al.2004.A&A}, as well as
2.4Ghz VLA radio data \citep{Bondi_et_al.2003.A&A}. To $K<20.5$, 766
objects from our catalogue have been observed spectroscopically in the
first epoch VVDS \citep{Lefevre_et_al.2004.astro.ph}.

Our primary objective in this paper is to describe in detail how our
near-infrared catalogues were prepared and to quantify their
reliability and completeness. Future papers will present more detailed
analysis of our $J-$ and $K-$selected samples and in particular the
clustering properties of objects with extreme colours.  All magnitudes
quoted in this paper are in the Vega system unless otherwise stated.

\section{Observations}\label{obs}

The survey described in this article covers in $J$ and $K$ bands four
different regions of the sky spanning a wide range of right ascension
so that at least one field is observable throughout the year. Each of
these regions has corresponding deep optical imaging, taken with
either the Canada France Hawaii Telescope's (CFHT) CFH12K camera
\citep{McCracken_et_al.2003.A&A,Lefevre_et_al.2004.A&A} or the ESO 2.2
metre telescope's Wide-Field Imager
\citep{Radovich_et_al.2004.A&A}. The observations described this paper
were carried out at the ESO New Techonology Telescope using the SOFI
Near Infrared imaging camera \citep{Moorwood1997Msgr} with the $J$ and
the $K_{s}$ filters. The $K_{s}$ filter is bluer and narrower than the
standard near-infrared $K-$band filter,  and so is less affected
by the thermal background of the atmosphere and of the telescope
\citep{Wainscoat_and_Cowie.AJ.1992}.  Throughout this paper our $K$
magnitudes are in fact ``$K_{s}$'' magnitudes. SOFI is equipped with a
Rockwell Hawaii HgCdTe 1024x1024 array and observations were made with
the Large Field (LF) objective, corresponding to a field-of-view of
$4.9\times4.9$ \arcsec and a pixels scale of $0.288$ \arcsec/pixel. 

\begin{figure}
\centering
\includegraphics[width=9cm]{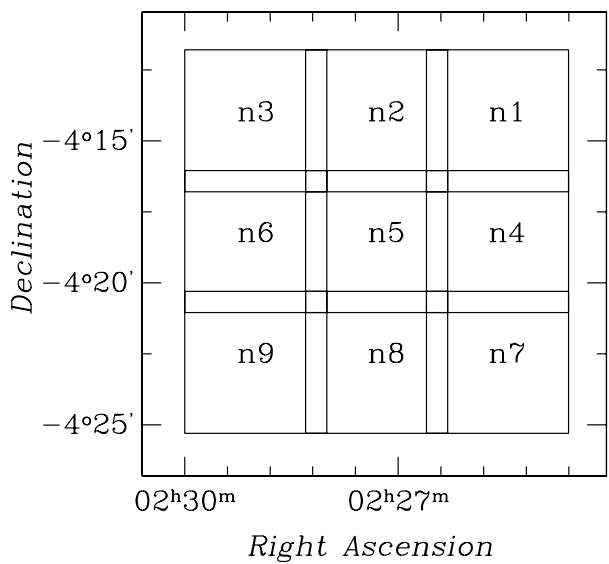} 

\caption{Layout of the observed $K-$band pointings for field
 $0226-04$.  The raster configuration chosen in order to secure a
 non-negligible overlap between adjacent pointings is clearly visible.
 Each pointing is is turn observed through a series of jittered
 exposures, with jitter box size of 30 \arcsec (see text for more
 details). }
\label{layout}
\end{figure}

The observations took place over a series of runs from September 1998
to November 2002. Each targetted field was observed in a series of
pointings in a raster configuration, each separated from surrounding
ones by $4'15$\arcsec in both right ascension and in declination, in
order to ensure a non negligible overlap between adjacent pointings.
Figure \ref{layout} shows the layout of our $K-$band observations of
field $0226-04$ as an example.

The well known peculiarities of infrared observations (higher and more
variable sky background, strong and variable absorption bands) dictated
our observational strategy. Total integration time per pointing was
around 1 hour for the $J$-band and three hours for $K-$band exposures,
with some pointings being observed for up to four hours in $K-$band.
Each integration consisted of many shorter jittered exposures with the
telescope being offset by random amounts within a box of size
30\arcsec. The jittered observations were usually grouped in
observation sequences each typically one hour long. Each individual
short exposure in both bands was 1.5 minutes long, with {\tt DIT = 15}
(meaning a detector integration time of 15 seconds was used) and {\tt
NDIT = 6} in $J$ (indicating that six of these 15 second integrations
were used) and with {\tt DIT = 10} sec and {\tt NDIT = 9} in $K$. For
the observations of the standard stars we adopted {\tt DIT = 2} ( {\tt
DIT = 1.2}) sec in $J$ (in $K$) and {\tt NDIT = 15} to avoid 
saturation.  \\

The four observed areas are listed in Table \ref{pointings} together
with the centres for each of the pointings (J2000), observing runs
when the observations were performed, total exposure times and
galactic extinction in $J-/K-$ band computed using the COBE/DIRBE dust
maps \citep{Schlegel_et_al.1998.ApJ}.  Unfortunately, poor weather
conditions partially hampered our efficiency and reduced our final
total areal coverage.

   \begin{table*}
\centering 
      \begin{tabular}{cccccccccc}
            \hline 
            \hline\noalign{\smallskip}
\multicolumn{10}{c}{ List of observed pointings}\\
            \hline\noalign{\smallskip}
field         & pointing & {\it RA} &{\it Dec}   &      & $J-$band &     &      & $K-$band   &    \\
            \hline\noalign{\smallskip}  
              &          &{\it(2000)}&{\it (2000)}& {\it Obs run} & $t_{exp}$ &{ $A_{J}$} & {\it Obs run} & $t_{exp}$ &{ $A_{K}$} \\
            \hline\noalign{\smallskip}  
$0226-04$     &  n1 & 02 27 14 & -04 14 18 & nov02  & 60 & 0.023&  sep98       &    195    & 0.010 \\
              &  n2 & 02 26 57 & -04 14 18 & sep00  & 55 & 0.024&  sep98       &    214.5  & 0.010 \\
              &  n3 & 02 26 40 & -04 14 18 & nov02  & 60 & 0.024&  sep98 \& nov98 & 217.5  & 0.010 \\
              &  n4 & 02 27 14 & -04 18 33 & nov02  & 66 & 0.023&  nov98       &    192    & 0.009 \\
              &  n5 & 02 26 57 & -04 18 33 & nov02  & 88 & 0.023&  nov98       &    180    & 0.009 \\
              &  n6 & 02 26 40 & -04 18 33 & sep00  & 60 & 0.024&  nov98       &    196.5  & 0.010 \\
              &  n7 & 02 27 14 & -04 22 48 & sep00  & 60 & 0.022&  nov98       &    180    & 0.009 \\
              &  n8 & 02 26 57 & -04 22 48 & nov02  & 60 & 0.023&  nov99       &    180    & 0.009 \\
              &  n9 & 02 26 40 & -04 22 48 & nov02  & 60 & 0.025&  nov99       &    180    & 0.010 \\
            \noalign{\smallskip}
            \hline
            \hline
$2217+00$     &  n1 & 22 18 08 & 00 19 45 & sep00 & 60  & 0.059 &  sep98       &    255  & 0.024 \\
              &  n2 & 22 17 51 & 00 19 45 & sep00 & 60  & 0.058 &  sep98       &    270  & 0.024 \\
              &  n3 & 22 17 34 & 00 19 45 & nov02 & 60  & 0.055 &  sep00 \& nov02 & 240  & 0.022 \\
              &  n4 & 22 18 08 & 00 15 30 & sep00 & 60  & 0.063 &  nov98 \& nov02 & 210  & 0.026 \\
              &  n5 & 22 17 51 & 00 15 30 & sep00 & 60  & 0.064 &        --       &  --  &   --  \\        
            \noalign{\smallskip}
            \hline
            \hline
$1003+01$     &  n1 & 10 03 56 & 01 58 54 & apr99    &    45      & 0.022  &  mar99    &    180  & 0.010 \\
              &  n2 & 10 03 39 & 01 58 54 & apr99    &    45      & 0.019  &  mar99    &    180  & 0.010 \\
              &  n3 & 10 03 22 & 01 58 54 & apr00    &    60      & 0.018  &  nov99    &    180  & 0.010 \\
              &  n4 & 10 03 56 & 01 54 39 & apr99 \& nov02 &  60  & 0.023  &  mar99 \& apr99 &  181.5 & 0.009 \\
              &  n5 & 10 03 39 & 01 54 39 & apr99 \& nov02 &  60  & 0.020  &  nov98 \& mar99 &  195   & 0.009 \\
              &  n6 & 10 03 22 & 01 54 39 &          --     &  -- &     -- &  apr00    &    180  & 0.010 \\
            \noalign{\smallskip}
            \hline
            \hline
$1400+05$ & n1 & 14 00 18 & 05 13 30  & apr00 &    60 & 0.022 & mar99          & 181.5 & 0.010 \\
          & n2 & 14 00 01 & 05 13 30  &  --   &   --  &  --   & mar99 \& apr00 &  222  & 0.010 \\
            \noalign{\smallskip}
            \hline
            \hline
            \noalign{\smallskip}

         \end{tabular}
\caption[ ]{List of fields observed. See text for more details.}
\label{pointings}
   \end{table*}

\section{Data reduction}

\subsection{Science frames} 

Data reduction of the scientific frames in both filters included the
usual standard steps: dark subtraction, flat fielding and sky
subtraction. We now outline the basic processing steps followed.  The
darks to be subtracted were computed using the IRAF\footnote{IRAF is
distributed by the National Optical Astronomy Observatories, which are
operated by the Association of Universities for Research in Astronomy,
Inc., under cooperative agreement with the National Science
Foundation.} task {\it darkcombine} from a series of darks obtained
with the same {\tt DIT} and {\tt NDIT} as the science frames.  The
well known complex bias behaviour of the Rockwell Hawaii array means
that there is a dependence of the detector dark on time and
illumination history, and this manifests itself as a pattern which
remains in all the images after dark subtraction.  This is visible as
a discontinuity between the lower upper part of the array and the
upper lower part of the array (\ie were the two upper quadrants join
the two lower quadrants). This pattern is a purely additive component
and it is removed in the sky subtraction step as it changes little in
images in the same stack of observations.

Flat-field frames were obtained for both bands with the {\tt ON-OFF}
procedure as described in the SOFI manual. We carefully checked whether
our flat-field frames contained the large jump between the lower and
upper part of the array which is a signature of a remaining dark
residual, rejecting those frames which had a non-negligible dark
residual. Using IRAF's {\it flatcombine} task we were able to derived
from the remaining flats the final $J$ or $K$ flat. To test the
accuracy of the ON-OFF flat field correction, we compared the counts
observed for each photometric standard star in different array
positions. The rms on the mean was always of the order of a few percent
for both bands (and usually below $4\%$).

Before proceeding with the data reduction, we checked the quality of
the images and rejected those with full-width half-maximum (FWHM)
values larger than $1.\arcsec3$ or those which were badly affected by
sky--transparency fluctuations based on the flux measurements of a
chosen reference star. The exposure times quoted in Table
\ref{pointings} are those obtained after this step.

We next used IRAF DIMSUM\footnote{DIMSUM is the Deep Infrared
  Mosaicing Software package, P. Eisenhard, M. Dickinson, S. A.
  Stanford, and J. Ward, available at
  ftp://iraf.noao.edu/iraf/contrib/dimsumV2/} package to subtract the
sky from the dark subtracted, flat fielded, science images
\citep{Stanford_et_al.1995.ApJ}. DIMSUM was used in a two step process
and on sets of 20-30 images belonging to the same observation sequence.

As a first step for each image 6 neighbouring images (3 minimum for the
images at the end/beginning) from the same observation sequence were
selected to obtain an initial sky estimate. The sky subtracted images
were then used to identify the image regions covered by the objects.
In this way for each image an object mask was defined and was used to
exclude those regions from a second, final pass of sky estimation and
subsequent subtraction. 

Finally, the second pass sky subtracted images of the same observation
sequence were combined in a stacked image by using the task {\it
dithercubemean} from the IRDR software
\citep{Sabbey_et_al.2001.adass}. This task uses bi-linear
interpolation to register the input frames, taking into account pixel
weights, based on image variance, exposure time and pixel gain (which
is calculated using superflat field image calculated from all the sky
subtracted images; see \citet{Sabbey_et_al.2001.adass} for more
details).

A final weight map was also produced by combining individual image
weights. At the end of these data reduction steps usually one or two
(three or four for $K-$band observations) coadded images and their
relative weight maps were produced for each pointing in $J-$band,  
depending on how the observations were sequenced.

\begin{figure}
\centering
\includegraphics[width=9cm]{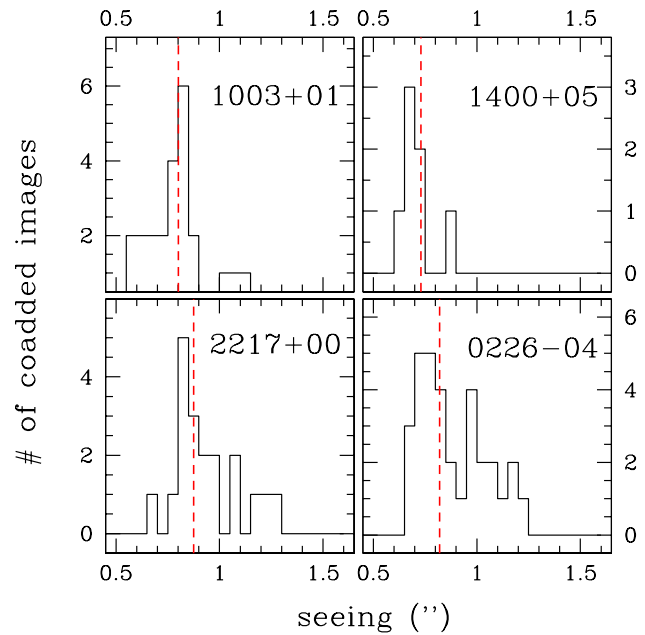}
\caption{FWHM seeing distribution of the coadded images for each
  field observed in $K-$band. The dashed line shows the median seeing
  measured in the final mosaicked image. Note that in all cases this
  value is below $1\arcsec$. All images with seeing higher than
  $1.\arcsec 3$ were discarded, and are not shown in this plot.}
\label{seeing}
\end{figure}

Figure \ref{seeing} shows the FWHM seeing distribution for the
different coadded $K-$band images for each of the four fields. Tables
\ref{maglimJ} and \ref{maglimK} list for each field and each band the
median FWHM seeing as measured in the final mosaicked image (for the
$K-$band images this value is shown by the dashed lines in
Figure~\ref{seeing}).

\subsection{Astrometric calibration} 

Astrometric calibration on the coadded images was performed in two
steps for all our images. We first computed a linear astrometric
solution using the astrometric catalogue of the United States Naval
Observatory (USNO)-A2.0 \citep{Monet.1998.AAS}, which provides the
positions of $0.5 \times 10^8$ sources. The area covered by each
coadded image usually contained around 10 objects (after removal of
saturated and extended sources). This first astrometric solution was
then improved by using a catalogue of sources extracted from the
resampled $I-$band images which is used as a reference catalogue for
all the other optical bands (see \citep{McCracken_et_al.2003.A&A}, and
the similar procedure adopted for the $U-$band in
\citet{Radovich_et_al.2004.A&A}).  As the surface density of these
objects is much higher than that of USNO-A2.0, a much higher accuracy
in the relative astrometric solution can be obtained. Such accuracy
allows us to match sources at the sub-pixel level between optical and
infra-red bands.

\begin{figure}
\centering
\includegraphics[width=9cm]{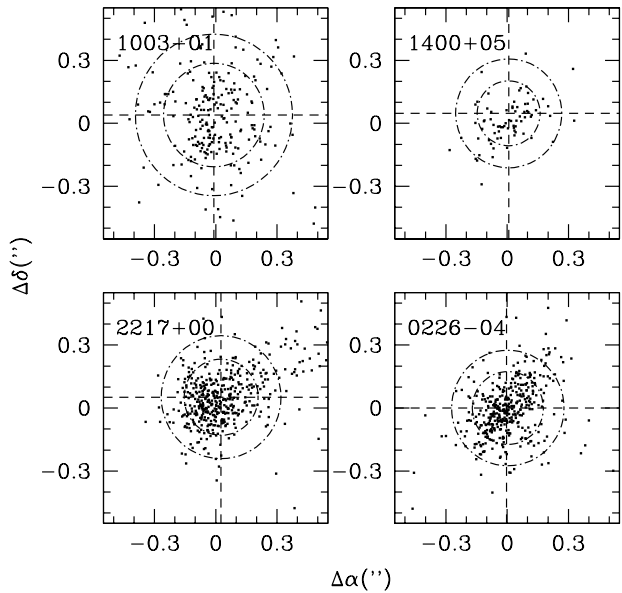} 
\caption{Radial residuals between $K-$band and $I-$band positions for
unsaturated, point-like sources for each of the four fields. The
inner and outer circles enclose 68\% and 90\% of all objects
respectively.  The dashed lines cross each other at the position of
the centroid of the residuals distribution. }
\label{astrometry}
\end{figure}

The quality of the astrometry in the final $K-$band mosaicked images
is shown in figure \ref{astrometry} for the four fields. Radial
residuals between $K-$band and $I-$band positions for unsaturated,
point-like sources are plotted for each of the four mosaicked fields.
The inner circle and the outer circle show in each field enclose 68\%
and 90\% of all objects, indicating that we have reached the level of
sub-pixel accuracy in the relative astrometry between $I-$band and
$K-$band images (the achieved RMS positional accuracy, defined as the
radius enclosing 68\% of the objects, is for all four fields around
$\sim 0.\arcsec 2$). Similar values of radial residuals are obtained
between $J-$band and $I-$band positions. The RMS accuracy of our
absolute astrometric solution is therefore of the same order of that
obtained for $I-$band data, that is $\sim 0.\arcsec 3$ (see
\citet{McCracken_et_al.2003.A&A}).

\subsection{Photometric calibration} 

The photometric calibration was performed using standard stars from
the list of near-infrared NICMOS standard stars
\citep{Persson_et_al.1998.AJ}.  At least two standard stars were
observed each night at low airmass ($sec(z)<1.3$), at intervals of
roughly two hours, in both $J$ and $K$ bands. Each standard star was
observed in 5 different array positions: once near the center of the
array and once in each of the four quadrants. These images were dark
subtracted and flat fielded according to the procedure discussed
above. Sky subtraction was obtained by subtracting the median of the
four adjacent images (usually the standard star fields are empty of
bright stars).

Instrumental aperture magnitudes for the standard stars were computed
within an $8\arcsec$ radius and were airmass corrected assuming an
atmospheric extinction coefficient of 0.1 (0.05) in magnitudes/airmass
in $J$ ($K$) bands (based on recent measurements in LaSilla, see SOFI
home page at ESO: www.ls.eso.org/lasilla/sciops/ntt/sofi/index.html).
We estimated the actual zero point for each night by comparing the
instrumental magnitudes with those quoted in the literature. This way
we obtained for each coadded image an absolute photometric calibration.

\begin{figure}
\centering
\includegraphics[width=9cm]{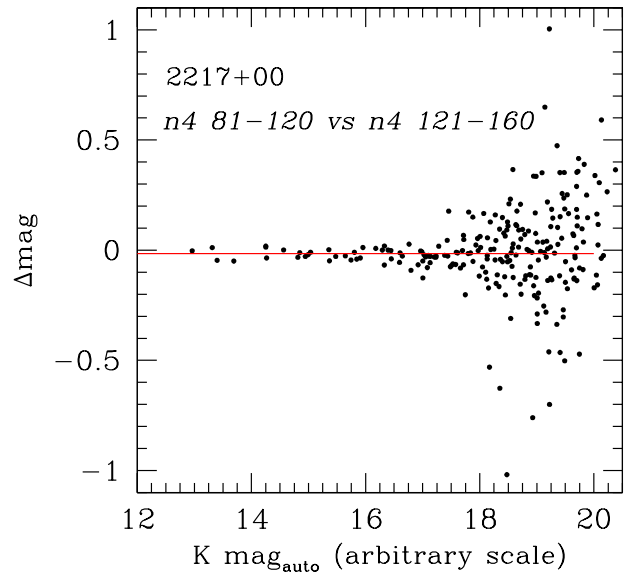}
\caption{ Comparison between the magnitude of corresponding objects in
two different $K-$band coadded images of the same pointing. The
sequence of images $ n4 ~ 81-120$ was observed during the run of
November 1998, while $ n4 ~ 121-160$ was observed during the run of
November 2002. The agreement for the photometry of these two coadded
images is quite good (the mean value of the magnitude difference is
only $\Delta mags = -0.015$).  }
\label{zeropoint_1}
\end{figure}

Not all our nights were of excellent photometric quality, and
therefore a refinement of this first photometric solution was needed.
After correcting each coadded image to one second exposure time, to
zero airmass and scaling to an arbitrary zero point (normally we chose
30) we improved on this first solution in two steps.

For different coadded images of the same pointing, when available, we
compared the magnitude of the point--like brighter objects in common
using {\tt SExtractor}'s \citep{Bertin_and_Arnouts.1996.A&AS} {\tt
MAG\_AUTO} measurements to perform the comparison.  In this way were
able to check for possible photometric offsets between different
coadded images of the same pointing. In case of discrepancies we
anchored the zero point to the coadded image with the best photometric
quality (based on the quality both of its night of observation and of
the specific sequence used to build it). Figure \ref{zeropoint_1}
shows as an example the comparison between two different coadded
$K-$band images of pointing n4 of field $2217+00$, the first taken
during the run of November 1998 and the second during the run of
November 2002. In this case the agreement between these two coadded
images is quite good. In other cases a shift had to be introduced in
order to put the different coadded images of the same pointing on the
same zero point scale.  The size of these shifts was in a few extreme
cases as large as 0.3 mags, but mostly below 0.1 mags.

As a second step, we used the stars in common in the overlapping areas
between different pointings to define a common photometric scale for
each field, assuring the homogeneity of the survey photometry. In this
case the few brightest non saturated stars in common enabled us to
define a common photometric solution for the different pointing
through scaling factors, to be used when building the final mosaicked
image of the field. The corresponding shifts introduced in the zero
point of the fields to be corrected were always below 0.1 mags.
 
\begin{figure}
\centering
\includegraphics[width=9cm]{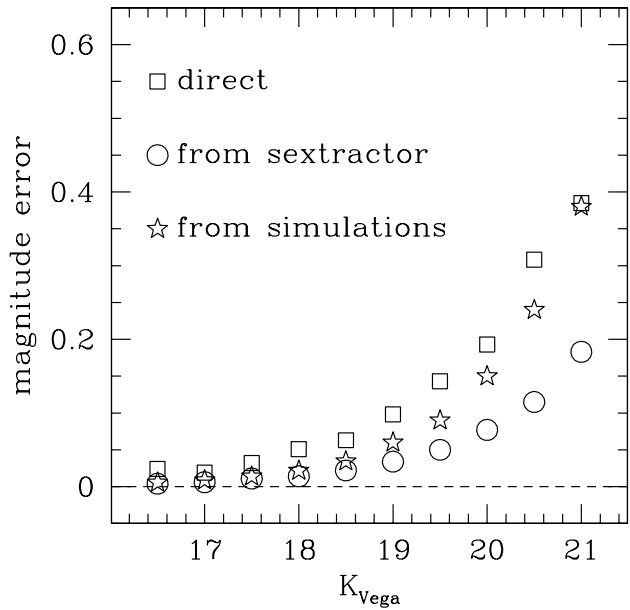} 
\caption{ Root-mean-square magnitude errors as a function of $K-$band
  magnitude. See text for more details. }
\label{magerrors}
\end{figure}

We took advantage of the repeated observations of the same pointing to
estimate the random photometric errors in the final mosaicked images.
Following the same procedure outlined in detail in the following
Sections, we produced for field $1400+05$ 3 mosaicked $K-$band images,
each corresponding to one hour of exposure time, and extracted a
photometric catalogue from each of these images. By comparing the
magnitudes measured for the same object in the three independent one
hour stacks, we obtained a direct estimate of the random photometric
errors present in our $K-$band data, including flat-fielding and/or
background subtraction inaccuracies (all the direct error estimate have
been divided by $\sqrt 3$ to take into account the shorter exposure
time for the individual stacks). Figure \ref{magerrors} shows the
comparison of such direct error estimate with the errors obtained by
{\tt SExtractor} and with the errors computed from the simulations used
in Sect.~\ref{CandC} to estimate completeness of our fields. The errors
obtained by {\tt SExtractor} are always lower, usually by a factor of
2, than the more realistic direct error estimate. The errors obtained
from simulations are slightly lower than the direct error estimate:
they refer to stellar objects, and are therefore less affected by
errors in background determination, especially at bright magnitudes. A
similar trend is observed for $J-$band magnitudes errors: the estimate
provided by {\tt SExtractor} is lower by a factor of roughly two than
the directly measured errors. A realistic estimate of random
photometric errors as a function of object magnitude has to be taken
into account when \eg using these data for photometric redshift
determination (see Bolzonella et al. 2005, in preparation).

\subsection{Preparation of final stacks} 

Once the astrometric and photometric solutions were computed for all
the coadded images of each field, these images, together with their
weight maps, were combined to produce the final stacks and their weight
maps. This is carried out in a two-step process by using Swarp, an
image resampling tool \citep{Bertin_et_al.2002.adass}. This process
does not differ from that described in detail for the stacking of
optical images described in \citet{McCracken_et_al.2003.A&A}, and
therefore the interested reader should check there for its details. The
final mosaicked images has the same pixel scale (0''.205/pixel) and
orientation as the final optical images
\citep{McCracken_et_al.2003.A&A}. For each field and band these images
were corrected for the mean galactic extinction at the centre of the
field as derived from the maps of \citet{Schlegel_et_al.1998.ApJ}.

\subsection{Catalogue preparation}

We used {\tt SExtractor} \citep{Bertin_and_Arnouts.1996.A&AS} to
extract sources from the stacked images and their weight maps. For
objects to be included in our $J-$ or $K-$band catalogues they must
contain at least 3 contiguous pixels above the detection threshold of
$1.1\sigma$, giving a minimum signal to noise ratio per source of
$1.9\sigma$ for this per-pixel threshold. This conservative threshold
means we minimise the number of spurious detections while not
adversely affecting our completeness (see later). For the mosaicked
image of field $0226-04$ a chi-squared $BVRIK$ image was constructed
\citep{Szalay_et_al.1999.AJ}, and for the $J-$ and $K-$band
catalogues, image detection was performed using such an image in {\tt
SExtractor} double--image detection mode, with similar extraction
parameters as those quoted above (the interested reader should refer
to \citep{McCracken_et_al.2003.A&A} for more details on the
chi-squared image construction). In our catalogues magnitudes were
measured using the {\tt SExtractor} parameter {\tt MAG\_AUTO}.  This
parameter is intended to give a precise estimate of total magnitudes
for extended objects, and is inspired by Kron's first moment algorithm
\citep{Kron.1980.ApJS}. We adopted a minimum Kron radius $R_{min} =
1''.5$, that is when objects are faint and unresolved {\tt MAG\_AUTO}
magnitudes revert to simple aperture magnitudes. Using the same
simulations adopted to test the completeness of our catalogues (see
Sect.~\ref{CandC} ) we verified that {\tt MAG\_AUTO} was a reliable
estimate of the input total magnitude of the simulated objects and
that the systematic loss of flux was always smaller, in the range of
magnitudes of interest, than the dispersion in the magnitudes
recovered. The catalogues were visually inspected and noisy border
regions of the mosaicked images were masked out together with circular
areas surrounding bright objects.

The total area of our survey, after bad regions are excised, is
$389$~arcmin$^2$ in $J-$band and $430$~arcmin$^2$ in $K-$band.  The
total number of detected sources is 6433 down to $J = 22.00$ (7823 down
to $J = 22.25$) and 8105 down to $K = 20.75$ (9539 down to $K = 21.00$).
Tables \ref{maglimJ} and \ref{maglimK} show for each field the final
area covered and the number of objects detected down to different
magnitude limits.

\subsection{Star-galaxy separation} \label{SGsep_sect} 

\begin{figure}
\centering
\includegraphics[width=9cm]{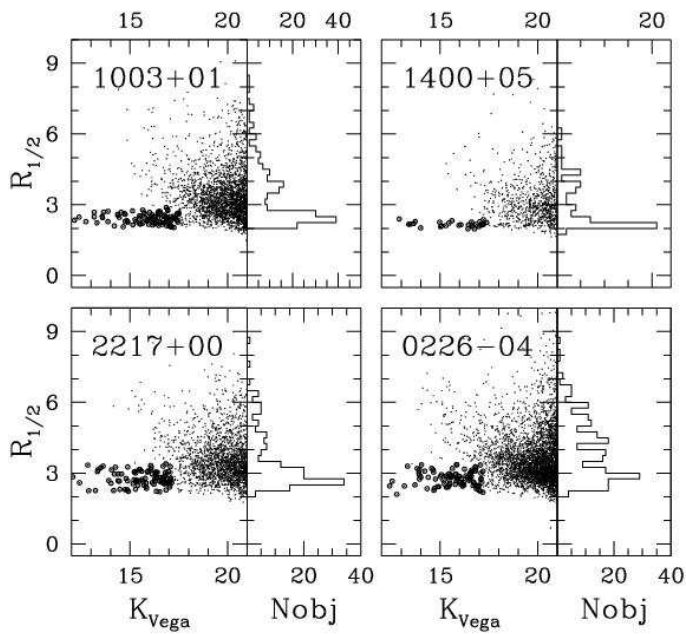}
\caption{Half-light radius ($r_{1/2}$) as a function of $K-$band 
  magnitudes for each of our four fields. Heavy dots show point like
  sources. The histogram on the right hand side shows clearly the locus
  of point-like sources for objects brighter than the classification
  limit. Fainter than this magnitude ($K\sim17$) the star-galaxy
  separation becomes unreliable.}
\label{sgsep}
\end{figure}

The separation of extended from point-like sources was performed
separately for each field and each band using the {\tt SExtractor}
{\tt flux\_radius} parameter. This parameter, denoted as $r_{1/2}$,
measures the radius that encloses 50\% of the object's total flux. For
point-like sources $r_{1/2}$ is independent of magnitude and depends
only of the image seeing.  As the seeing is quite uniform across our
mosaicked images for each field, a plot of $r_{1/2}$ vs. $J-$, or
$K-$band magnitude clearly defines the stellar locus. Heavy dots in
figure \ref{sgsep} show, for each field in $K-$band, the point-like
objects selected using this classifier. The histogram on the right
hand side of each panel illustrates for each field the distribution of
$r_{1/2}$ for magnitudes brighter than those deemed feasible for a
reliable star-galaxy separation. The stellar locus is clearly visible
in this plot.  A similar plot was also used to select stars in the
$J-$band images.

\section{Data quality assessment} 

In these subsections we present a series of quality assessment tests 
carried out on the catalogues prepared in the previous sections. 

\subsection{Completeness and contamination} \label{CandC} 

A simple estimate of our limiting magnitude can be obtained by using
the background RMS $\sigma$ provided by {\tt SExtractor} to compute,
for each stacked image, the corresponding $3\sigma-$ ($5\sigma-$)
magnitude limits.  The formula is $mag(n\sigma) = z_{p}
-2.5log(n\sigma\sqrt A)$, were n=3(5), $z_{p}$ is the zero point and
$A$ is the area of an aperture whose radius is the average FWHM of
point like sources (see Tables \ref{maglimJ} and \ref{maglimK}). The
values estimated this way do not vary much from field to field, and
are $mag (3\sigma)\sim 23.4$ and $mag (5\sigma)\sim 22.9$ for the
$J-$band images, while are $mag (3\sigma)\sim 22.6$ and $mag
(5\sigma)\sim 22.1$ for the $K-$band images. These values can be
regarded as indicative lower limits on the detectability of objects in
our catalogues.

\begin{figure}
\centering
\includegraphics[width=9cm]{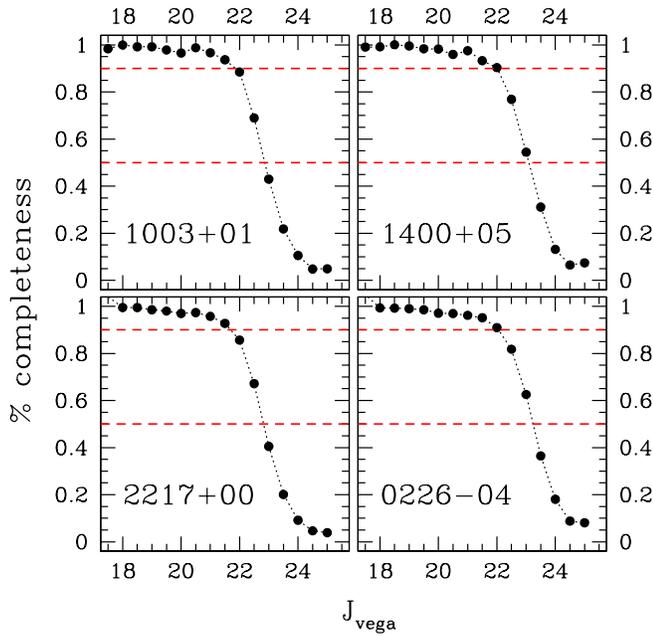} 
\caption{Ratio of detected input sources as a function of input
  magnitude for $J-$band. The input catalogue consists of a flat
  distribution of simulated point-like sources. The dotted lines show
  the adopted 50\% and 90\% completeness magnitudes for each field.  }
\label{incomp_J}
\end{figure}

\begin{figure}
\centering
\includegraphics[width=9cm]{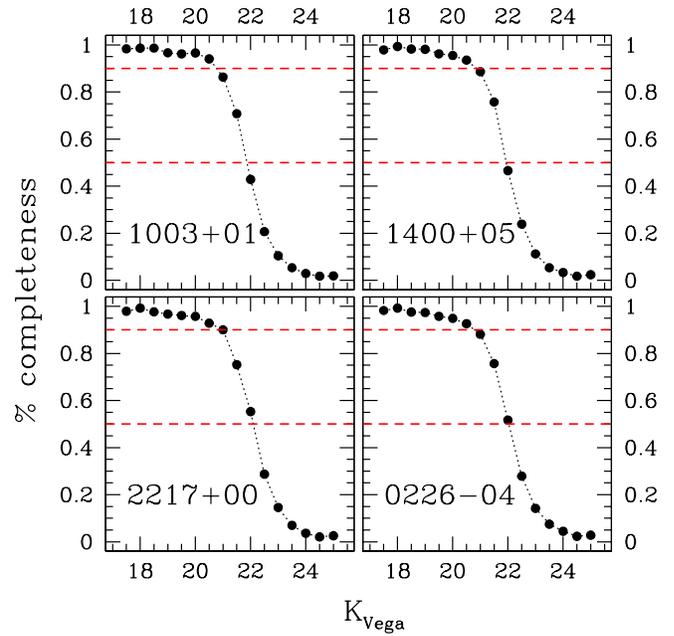} 
\caption{Ratio of detected input sources as a function of input
magnitude for $K-$band. The input catalogue consists of a flat
distribution of simulated point-like sources. The dotted lines show
the adopted 50\% and 90\% completeness magnitudes for each field. } 
\label{incomp_K}
\end{figure}

To better characterise the photometric properties of our images we
carried out an extensive set of simulations. For each mosaicked image
and for each filter a list of random positions was generated and
cross-checked with the position of the real sources to reject cases of
possible overlapping between generated random positions and bright
real sources positions. The remaining list of coordinates was used to
add to the stacks artificial stars distributed uniformly in the range
$17.5 < J/K < 25$.  {\tt SExtractor} was then run on these images with
the same parameters adopted for the detection of real objects and the
resulting catalogues were cross-correlated with the input list of
artificial stars. The process was repeated until a robust statistic
was obtained (more than 5000 objects per half-magnitude bin) and the
ratio $n_{\rm{input}}/n_{\rm{output}}$ was estimated in bins on
0.5~mags. The results are plotted in Figures \ref{incomp_J} and
\ref{incomp_K} for each field, and the values $mag_{\rm{comp}}$ and
$mag_{\rm{lim}}$, corresponding respectively to $90\%$ and $50\%$
completeness level, are shown in Tables \ref{maglimJ} and
\ref{maglimK}.  Such completeness limits and curves assume that the
profile of the source is point--like and therefore should be
considered only as upper limits. The true limiting magnitude of an
extended source will depend on its light--profile and actual size and
can be up to one magnitude brighter for low surface brightness
objects, see eg.  \citet{Cristobal_et_al.2003.ApJ}.  For field
$0226-04$, where the extraction was done on the BVRIK chi-squared
image, the values quoted for completeness, being obtained from the $J$
and $K-$band mosaicked images only, are a conservative estimate,
irrespective of color, for the detectability of objects in the
chi-squared image. For each field we also measured incompleteness as a
function of position across the mosaicked image. Such tests did not
show any significant variation at $m = mag_{\rm{comp}}$ across the
mosaicked images (except for the small areas around bright stars,
those masked out in the final catalogue).

To complete the characterisation of the photometric properties of our
fields we estimated contamination rates as a function of magnitude
bins. Each mosaicked image was multiplied by $-1$ and detection of
(obviously spurious) objects was performed on this inverse image using
the same parameters adopted for detecting objects on the mosaicked
image.  For the field F02, where the detection was carried out using
the chi-squared image, we used the same technique described in
\citet{McCracken_et_al.2003.A&A}, that is the detection of fake objects
was performed on the inverse image using the chi-squared BVRIK image as
the reference image.  Tables \ref{maglimJ} and \ref{maglimK} list, for
each field and band, $m_{\rm{cont} < 10\%}$: the center of the faintest
0.5~mags bin where contamination by spurious sources is below 10\%.
For the magnitude limit for the scientific analysis of each of our
fields we adopted the conservative choice of the brightest between
$m_{\rm{comp}}$ and $m_{\rm{cont} < 10\%}$. As expected, field
$0226-04$ has the lowest contamination rate at the completeness limits
of the survey, confirming the effectiveness of the chi-squared
detection image technique in reducing the number of spurious
detections.

\begin{table*}
      \begin{flushleft}
      \begin{tabular}{ccccccccc}
      \hline field name & seeing & $m_{comp}$ & $m_{lim}$ & $m_{cont <
10\%}$ & Area & $N_{TOT}$ & N($m<22.00$) & N($m<22.25$) \\ & arcsec&
& & & arcmin$^2$ & & & \\ \hline\noalign{\smallskip}
$0226-04$ & 1.05 & 22.25  & 23.25 & 23.00 & 162.6 & 12327 & 2604 &  3144 \\
            \hline
$2217+00$ & 1.00 & 21.90  & 22.90 & 21.75 & 102.5 & 6676  & 1659 &  1949 \\
            \hline
$1003+01$ & 0.90 & 22.25  & 22.90 & 21.50 & 102.7 & 8072  & 1689 &  2125 \\
            \hline
$1400+05$ & 1.20 & 22.00  & 23.15 & 21.90 & 23.4  & 1589  & 336  &   408 \\
            \hline
            \noalign{\smallskip}

         \end{tabular}
      \end{flushleft}
\caption[ ]{For each field for $J-$band data this table lists
$mag_{comp}$, \ie the center of the 0.5~mag bin where 90\% of
point--like sources are retrieved, $mag_{\rm{lim}}$ \ie the center of
the 0.5~mag bin where 50\% of sources are retrieved, and $m_{\rm{cont}
< 10\%}$, \ie is the center of the 0.5~mag bin where contamination by
spurious sources is below 10\%. See text for more details.}
\label{maglimJ}
\end{table*}

\begin{table*}
       \begin{flushleft}
      \begin{tabular}{ccccccccc}
            \hline 
field name & seeing & $m_{comp}$ & $m_{lim}$ & $m_{cont < 10\%}$ & Area & $N_{TOT}$ & N($m<20.75$) & N($m<21.00$) \\ 
                  & arcsec &            &           &                   &  arcmin$^2$ & & & \\ 
            \hline\noalign{\smallskip}
$0226-04$ & 0.82 & 20.75  & 22.00 & 21.50 & 168.3 & 12755 & 3161 & 3757 \\
            \hline
$2217+00$ & 0.87 & 21.00  & 22.15 & 20.75 & 88.3  & 5850  & 1838 & 2113 \\
            \hline
$1003+01$ & 0.80 & 20.75  & 21.90 & 20.50 & 122.0 & 9114  & 2235 & 2654 \\
            \hline
$1400+05$ & 0.73 & 20.75  & 21.90 & 20.50 & 45.0  & 3486  & 733  &  870 \\
            \hline
            \noalign{\smallskip}
         \end{tabular}
      \end{flushleft}
\caption[ ]{As for Table \ref{maglimJ} but for $K-$band data.  See
text for more details.}
\label{maglimK}
\end{table*}

\subsection{Galaxy and star number counts}

\begin{figure}
\centering
\includegraphics[width=9cm]{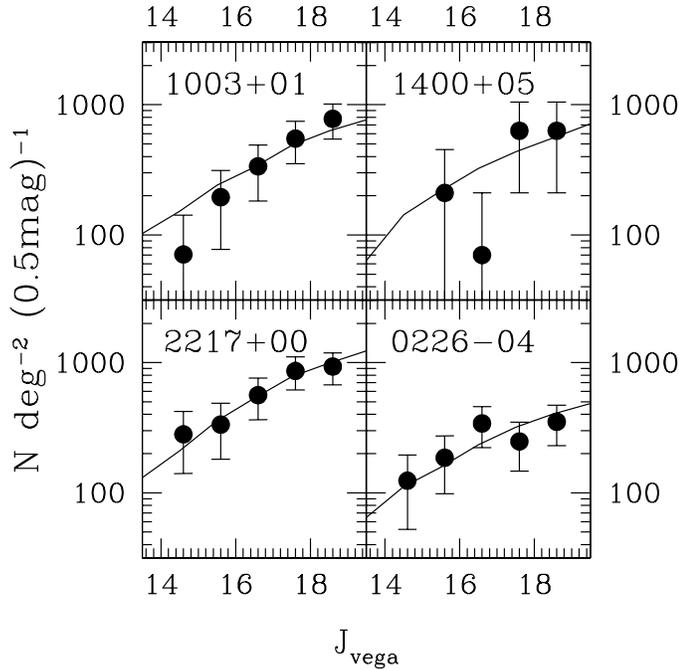} 
\caption{Differential number counts of stars in $J-$band for our
  fields. The continuous lines are the prediction of the model of
  \citet{Robin_et_al.2003.A&A} for each field.}
\label{star_counts_fields_J}
\end{figure}

\begin{figure}
\centering
\includegraphics[width=9cm]{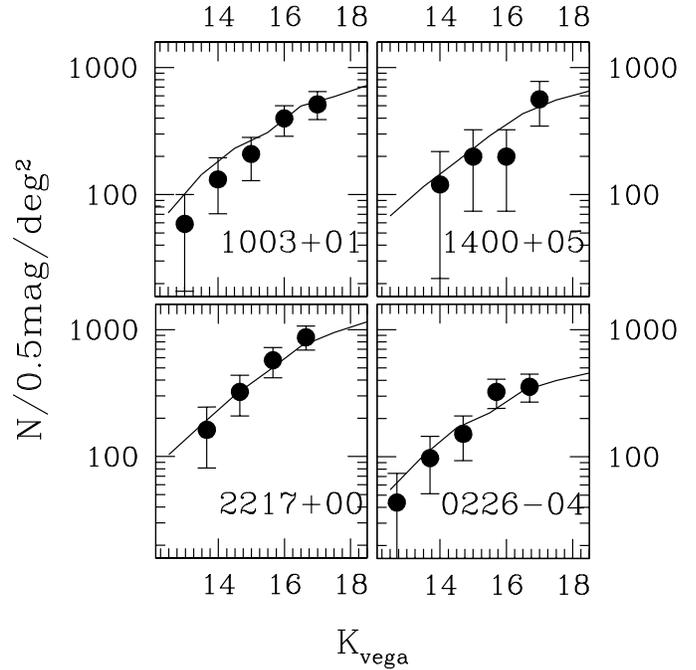} 
\caption{Differential number counts of stars in $K-$band for our
fields. The continuous lines are the prediction of the model of
\citet{Robin_et_al.2003.A&A} for each field.}
\label{star_counts_fields_K}
\end{figure}

Comparing number counts of galaxies and stars with published
compilations is a good check both of the star-galaxy separation
efficiency and of the reliability of our photometry, as well as the
sample reliability and completeness. The differential number counts of
stars (number 0.5~mag$^{-1}$ deg$^{-2}$) for each of our fields are
shown in Figures \ref{star_counts_fields_J} and
\ref{star_counts_fields_K}.  To avoid underestimating bright stars
counts, for this exercise we used the catalogues before excising the
areas around bright stars. The continuous lines are the prediction of
the model of \citet{Robin_et_al.2003.A&A} computed at the galactic
latitude appropriate for each field. The agreement between observed
and predicted star counts is very good for both $J$ and $K$ bands (the
error bars shown are Poissonian error bars), confirming the
reliability both of our photometry and of our star-galaxy separation
procedure. It should be noted that for field $2217+00$ the expected
star counts are quite high, due to the relatively low galactic
latitude of this field.  For fields $1003+01$ and $1400+05$ the star
counts are slightly lower, while for $0226-04$ we can safely assume
that the star contamination is well below a few percent for all the
relevant magnitude bins (see later).

\begin{figure}
\centering
\includegraphics[width=9cm]{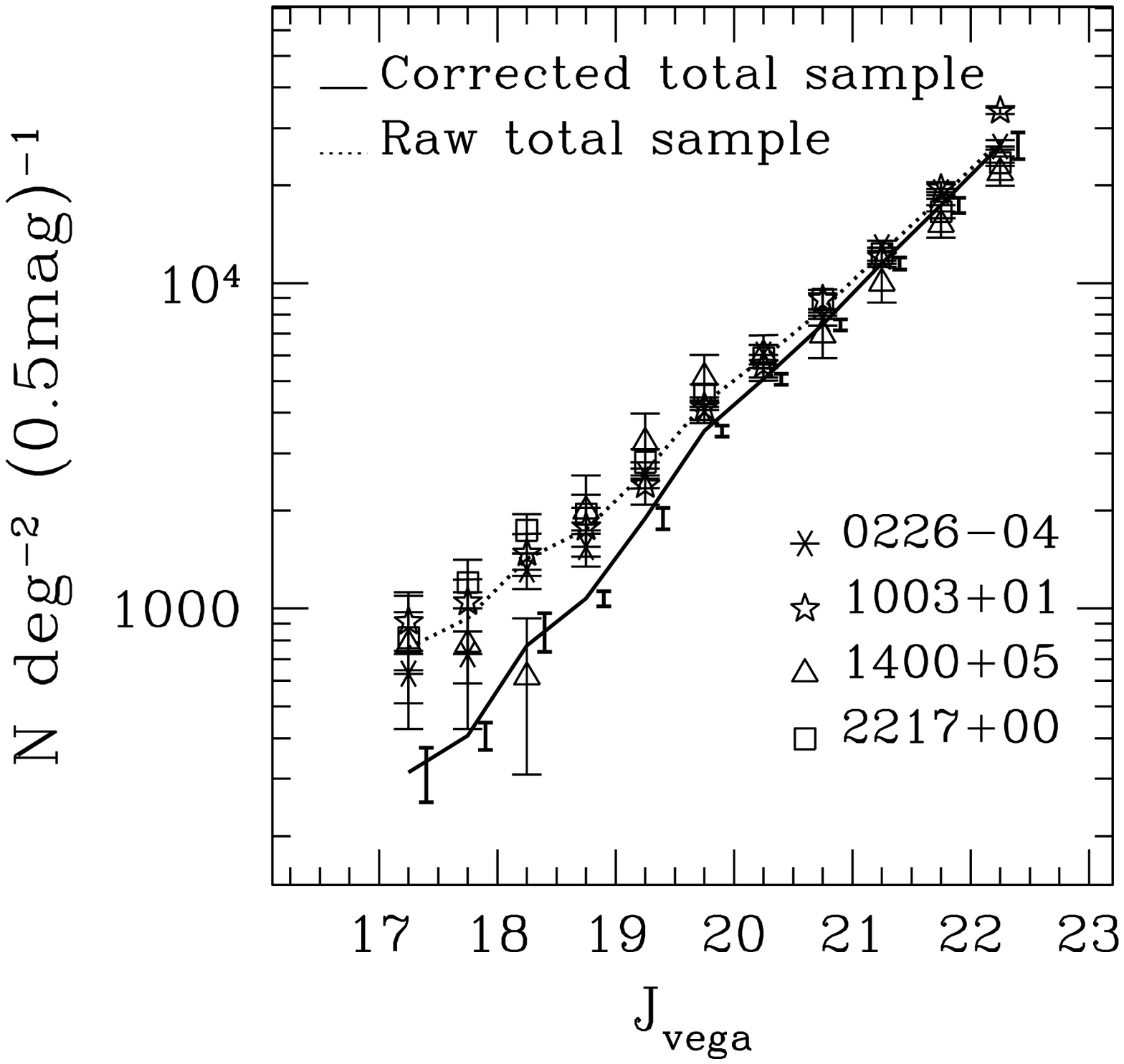} 
\caption{Differential $J-$band number counts in our four fields. Error
bars are Poissonian and no correction for stellar contamination has
been applied to the points plotted. The dotted line shows the final
total raw number densities, obtained by simply adding up raw densities
from the four fields, while the heavy line shows the total
galaxy densities obtained after correcting for star contamination. On
the right of each points, slightly offset for sake of clarity, are
shown $+/-1~\sigma$ error bars for total galaxy densities in each
bin. See text for more details.  }
\label{gal_counts_fields_J}
\end{figure}

\begin{figure}
\centering
\includegraphics[width=9cm]{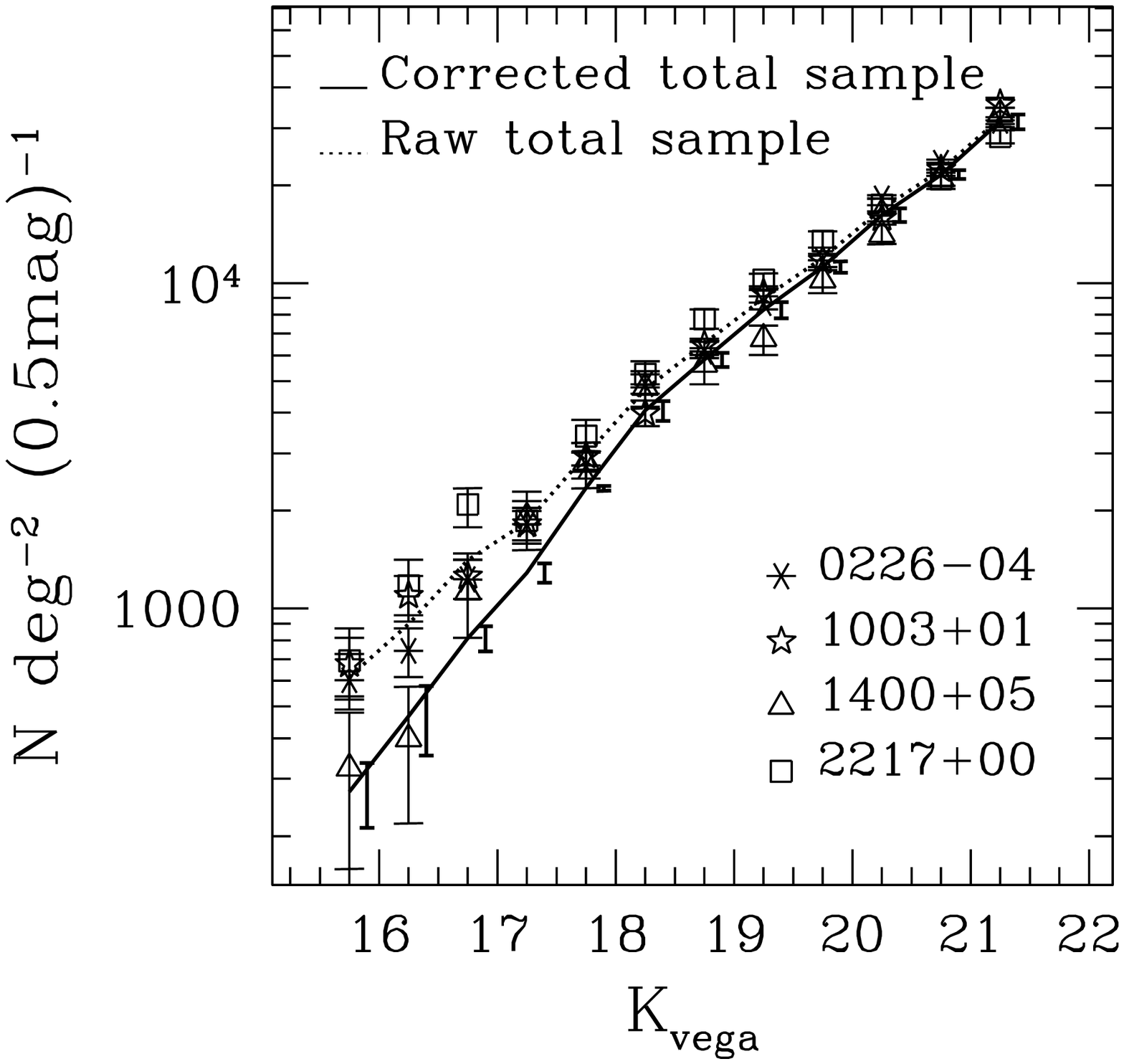} 
\caption{As in Figure \ref{gal_counts_fields_J}, but for $K-$band.
}
\label{gal_counts_fields_K}
\end{figure}

Figures \ref{gal_counts_fields_J} and \ref{gal_counts_fields_K} show
the differential number counts (number~(0.5~mags)$^{-1}$~deg$^{-2}$)
for our four fields in $J$ and $K$ bands, obtained by normalising the
observed raw number counts to the areas listed in Tables \ref{maglimJ}
and \ref{maglimK}. The error bars shown are Poissonian error bars and
no correction for stellar contamination has been applied the counts
shown. For the $0226-04$ field the contamination estimated from the
prediction of the model of \citet{Robin_et_al.2003.A&A} is below 5\%
for the fainter bins shown in the plot, while for the brighter ones
(below $J= 19.5$) rises beyond 10\%.  The situation is not so
favorable for fields $1400+05$ and $1003+01$: for these two fields
only at magnitudes fainter than $J = 20.5$ contamination rates falls
well below 10\%.  The worst case is field $2217+00$ where, due to the
lower galactic latitude of this field only at magnitudes fainter than
$J=21.0$ do contamination rates become negligible. A similar trend
holds for the $K-$band counts. Having taken into account these
effects, the agreement among the fields is quite remarkable. The
dotted line shows, both in Figure \ref{gal_counts_fields_J} and in
Figure \ref{gal_counts_fields_K}, the final raw counts
(number~0.5mag$^{-1}$ deg$^{-2}$), obtained by simply adding up counts
from the four fields, while the heavy continuous line shows the total
galaxy counts obtained after correcting each field counts for stellar
contamination according to the predictions of the model of
\citet{Robin_et_al.2003.A&A}. Given the excellent agreement between
the model of \citet{Robin_et_al.2003.A&A} and our bright star counts
(see Figures \ref{star_counts_fields_J} and
\ref{star_counts_fields_K}), we always use this model to correct for
stellar contamination, even in the brightest magnitude bins. On the
right of the continuous line, slightly offset for sake of clarity, are
shown $+/-1~\sigma$ error bars for total galaxy counts in each bin,
obtained by computing the weighted variance among corrected galaxy
counts for each field (with weighting proportional to area
covered). The size of these error bars indicates that taking into
account the different areas covered for each field, and after
correcting for stellar contamination, results from each field are
consistent at roughly the 5\% level. We did not need to apply any
other completeness or contamination correction to our data because
down to the limiting magnitude plotted in Figures
\ref{gal_counts_fields_J} and \ref{gal_counts_fields_K} such
corrections are negligible.

Tables \ref{counts_J} and \ref{counts_K} list our raw differential
number counts in each field, the total raw number densities (in units
of number~0.5~mag$^{-1}$ deg$^{-2}$), and the final, corrected for
stellar contamination, galaxy densities together with their $1\sigma$
error bars, computed as described above, for our total sample.

\begin{table*}
\begin{flushleft}
  \begin{tabular}{cccccccc}
\hline
$J_{\rm{vega}}$ & $0226-04$ & $2217+00$& $1003+01$ & $1400+05$ & {\it N$_{raw}$} & {\it N$_{corr}$}& $ \sigma $ \\
           & raw  & raw      & raw     & raw     & $0.5~$mag$^{-1}\rm{deg}^{-2}$ & $0.5~$mag$^{-1}\rm{deg}^{-2}$ & \\
\hline
17.25  &    28    &     23   &     26  &       5 &     758  &     311 &     63  \\
17.75  &    32    &     34   &     30  &       5 &     934  &     399 &     36  \\
18.25  &    59    &     49   &     42  &       4 &    1424  &     770 &    116  \\
18.75  &    69    &     56   &     51  &      13 &    1748  &    1087 &     49  \\
19.25  &   115    &     80   &     68  &      21 &    2627  &    1889 &    143  \\
19.75  &   183    &    129   &    118  &      33 &    4283  &    3510 &    133  \\
20.25  &   269    &    169   &    160  &      39 &    5892  &    5056 &    201  \\
20.75  &   347    &    254   &    253  &      45 &    8316  &    7462 &    293  \\
21.25  &   581    &    350   &    345  &      65 &   12404  &   11534 &    484  \\
21.75  &   849    &    470   &    557  &      99 &   18268  &   17365 &    960  \\
22.25  &  1183    &    673   &    969  &     141 &   27435  &   26516 &   2500  \\
            \hline
            \noalign{\smallskip}

          \end{tabular}
        \end{flushleft}
        \caption{Columns 2 to 5 show, for each of our fields, the raw
differential number counts per half magnitude bin. Column 6 show our
raw total number densities (number~0.5mag$^{-1}$ deg$^{-2}$). Column 7
shows our total, differential galaxy densities, corrected for star
contamination and Column 8 is their $1\sigma$ error bar, obtained from
the variance of the counts among the different fields.} 
\label{counts_J}
\end{table*}

\begin{table*}
\begin{flushleft}
  \begin{tabular}{cccccccc}
\hline 
$K_{\rm{vega}}$ & $0226-04$ & $2217+00$& $1003+01$ & $1400+05$ &{\it N$_{raw}$} & {\it N$_{corr}$}& $ \sigma$ \\
           & raw      & raw     & raw      & raw      & $0.5$~mag$^{-1}\rm{deg}^{-2}$ & $0.5~$mag$^{-1}\rm{deg}^{-2}$ & \\
\hline 
15.75  &     28   &      17   &     23   &       4 &    615     &     274  &      62  \\
16.25  &     34   &      29   &     37   &       5 &    897     &     467  &     113  \\
16.75  &     57   &      51   &     43   &      14 &   1409     &     907  &     136  \\
17.25  &     84   &      46   &     62   &      24 &   1845     &    1287  &      89  \\
17.75  &    128   &      84   &    100   &      35 &   2963     &    2343  &      42  \\
18.25  &    228   &     129   &    136   &      60 &   4722     &    4062  &     284  \\
18.75  &    287   &     191   &    217   &      70 &   6532     &    5823  &     284  \\
19.25  &    406   &     250   &    313   &      84 &   8992     &    8275  &     469  \\
19.75  &    545   &     333   &    394   &     128 &  11955     &   11215  &     453  \\
20.25  &    836   &     424   &    543   &     178 &  16916     &   16189  &     780  \\
20.75  &   1077   &     516   &    758   &     261 &  22304     &   21596  &     713  \\
            \hline
            \noalign{\smallskip}
            
          \end{tabular}
        \end{flushleft}
        \caption{As in Table \ref{counts_J}, but for $K-$band
counts.}
\label{counts_K}
\end{table*}

\begin{figure}
\centering
\includegraphics[width=9cm]{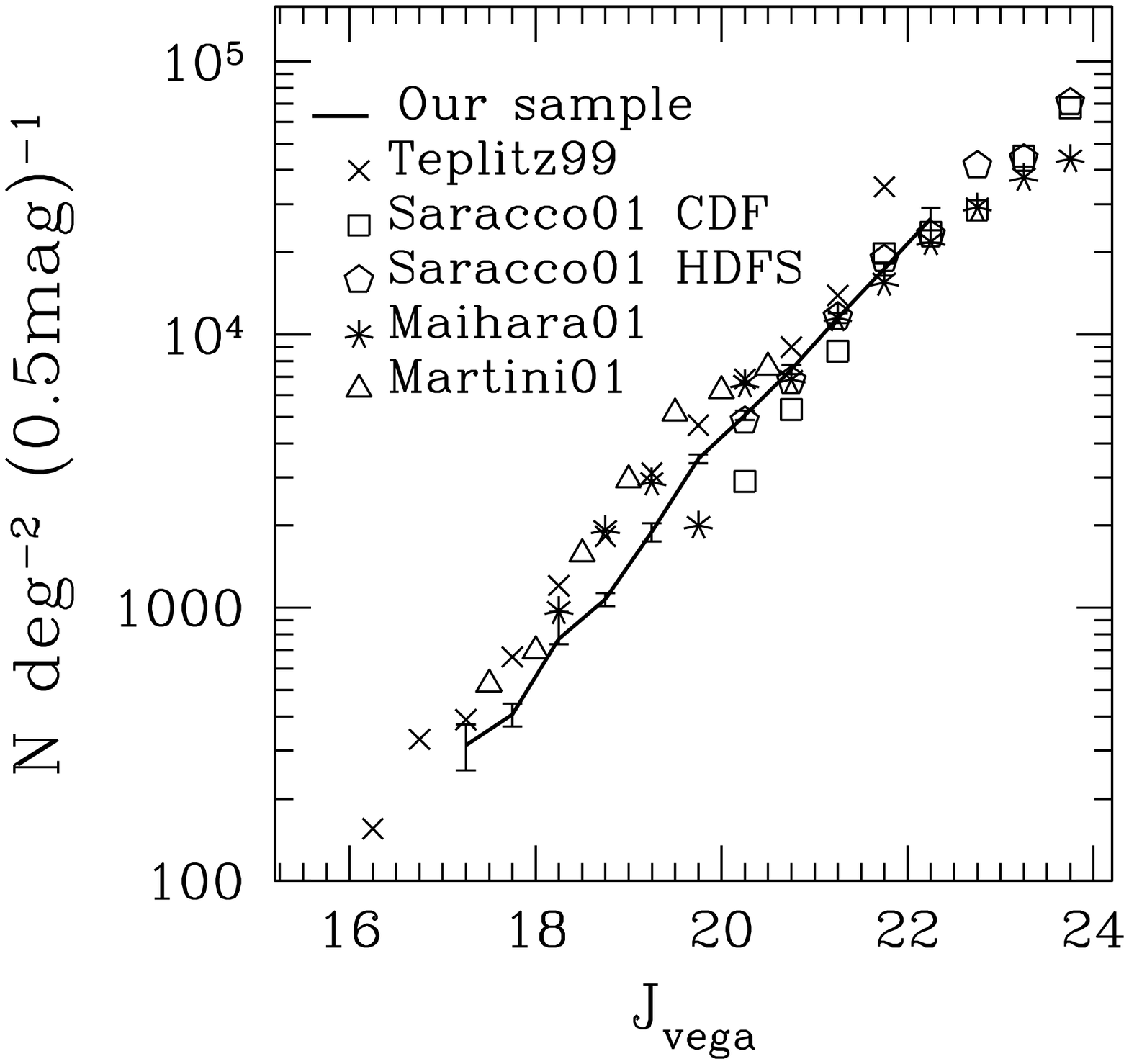} 
\caption{Our $J-$band galaxy number counts, in units of
  number~0.5mag$^{-1}$ deg$^{-2}$ compared to a literature compilation, including counts from
  \citet{Teplitz_et_al.1999.ApJ}, \citet{Saracco_et_al.2001.A&A},
  \citet{Maihara_et_Al.2001.PASJ} and \citet{Martini.2001a.AJ}.  }
\label{gal_counts_literature_J}
\end{figure}

\begin{figure}
\centering
\includegraphics[width=9cm]{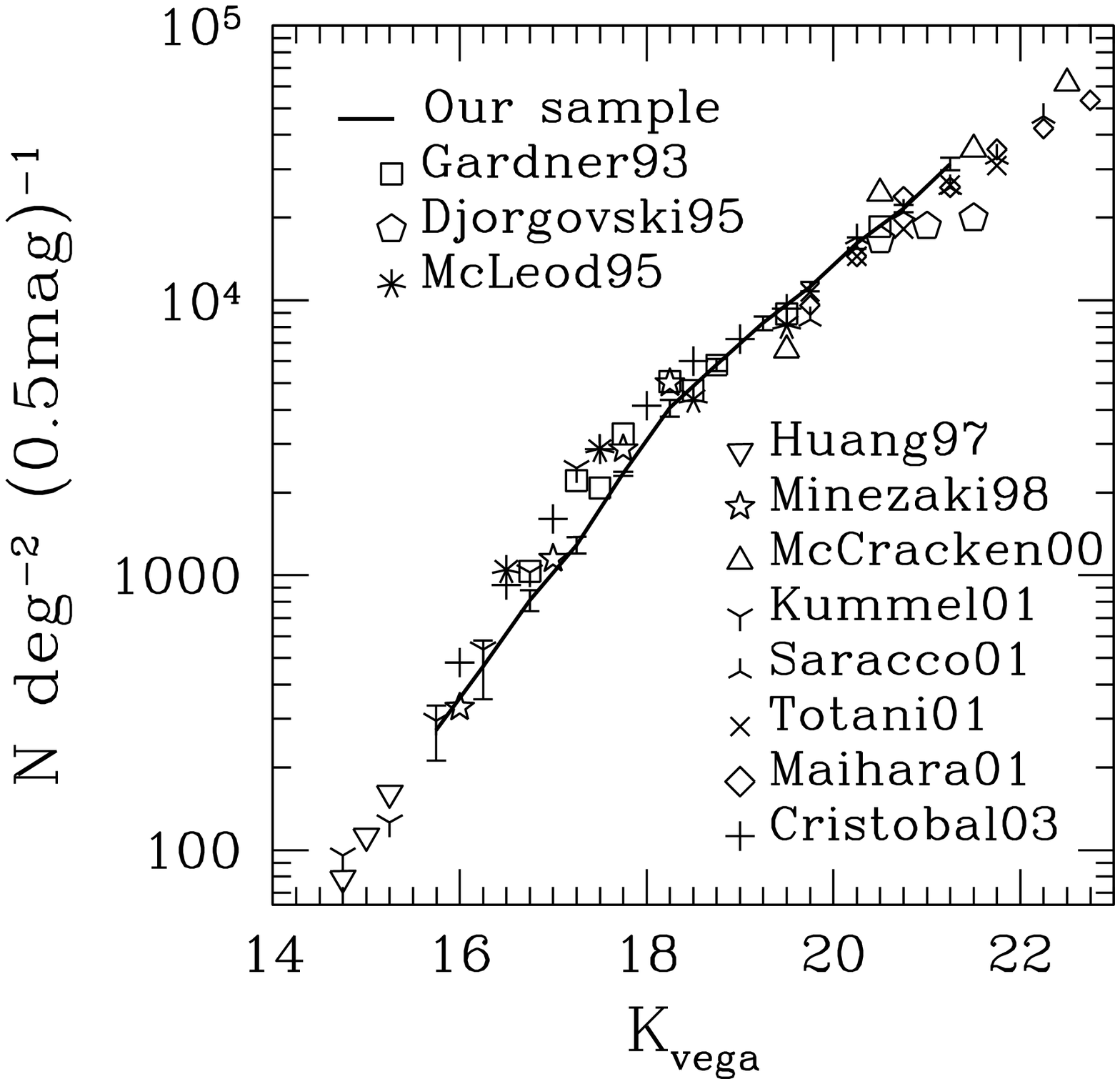} 
\caption{The $K-$band galaxy number counts, in units of
number~0.5mag$^{-1}$ deg$^{-2}$, obtained in this paper are compared
to a compilation of those from the literature, including counts from
\citet{Gardner_et_al.1996.MNRAS}, \citet{Djorgovski_et_al.1995.ApJ},
\citet{McLeod_et_al.1995.ApJS}, \citet{Huang_et_al.2001.A&A},
\citet{Minezaki_et_al.1998.ApJ}, \citet{McCracken_et_al.2000b.MNRAS}, 
\citet{Kummel_and_Wagner.2001A&A}, \citet{Saracco_et_al.2001.A&A},
\citet{Totani_et_al.2001ApJ}, \citet{Maihara_et_Al.2001.PASJ} and 
\citet{Cristobal_et_al.2003.ApJ}.
}

\label{gal_counts_literature_K}
\end{figure}

Figures \ref{gal_counts_literature_J} and \ref{gal_counts_literature_K}
show our total corrected galaxy counts (solid line) compared with a
selection of literature data. In the case of the $K-$band counts we
have followed the approach of \citet{Cristobal_et_al.2003.ApJ} and
select only reliable counts data from the literature, considering only
data with negligible incompleteness correction and with star-galaxy
separation applied. Given the relatively small number of published
$J-$band counts we decided to plot most of the available data. It
should be noted that we have been conservative in the selection of the
magnitudes interval plotted in our counts, restricting ourselves to
bins with relatively large numbers of galaxies, negligible
incompleteness and small contamination corrections. The agreement with
literature data is very good. For $J-$band data the estimated slope of
the galaxy counts in the range $17.25 < J < 22.25$, using a
weighted least-squares fit, is $\gamma_{J} \sim 0.39 \pm 0.06$,
consistent with the findings of e.g. \citet{Saracco_et_al.2001.A&A}.
The $K-$band galaxy counts show an evident change of slope around
$K \sim 18.0$. In the range $18 < K < 21.25$ the slope of
the galaxy counts $\gamma_{K} \sim 0.29 \pm 0.08$, with no significant
hints of steeper slope down to the faintest magnitude levels. In the
brighter magnitude range, $15.75 < K < 18$, the slope is
steeper: $\gamma_{K} \sim 0.47 \pm 0.23$.  Both results are consistent
with the findings of \citet{Gardner_et_al.1996.MNRAS} and
\citet{Cristobal_et_al.2003.ApJ}, who also find a similar break,
although at a slightly brighter magnitudes ($K \sim 17.5$).

\subsection{Star and galaxy colours}

\begin{figure}
\centering
\includegraphics[width=8.5cm]{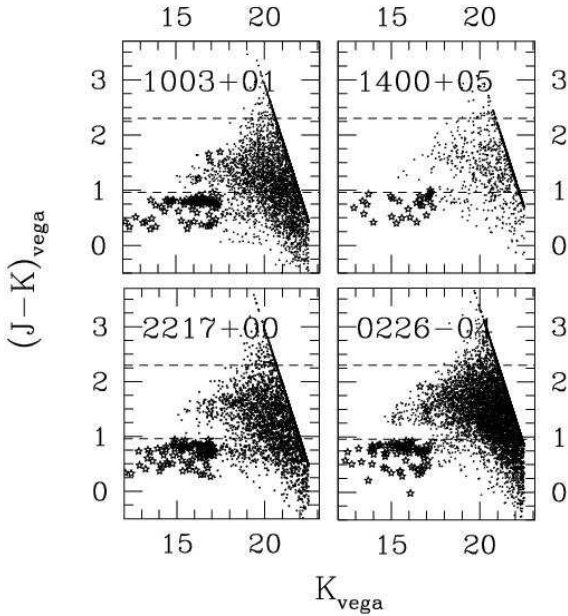}
\caption{Colour-magnitude diagrams for the $K-$selected samples in each
  of our fields. The star symbols indicate objects classified as
  point--like using our star/galaxy classifier. See text for more
  details.}
\label{JmKvsK}
\end{figure}

In this section we will further evaluate the quality of our absolute
and relative photometric calibration by investigating the colors of
stars/galaxies in our fields. As the $K-$band data have a slightly
better seeing than $J-$band data (see Tables \ref{maglimJ} and
\ref{maglimK}), to perform this analysis we used the sample of stars
defined using Figure \ref{sgsep}, 282 in total. We used {\tt
SExtractor} in dual-image mode to measure colours using matched
apertures, using $K-$band images for detection and {\tt mag\_{AUTO}},
based on the $K-$band flux distribution, to measure magnitudes in $K$
and $J$ bands. For the mosaicked images of field $0226-04$, as usual,
the chi-squared BVRIK was used as reference image for photometric
measurements, while stars were selected based on their $K-$band image
parameters. We checked that the difference between colors obtained
using {\tt MAG\_AUTO} are consistent with those one would obtain using
a classical aperture magnitude. Using only the good data quality area
common to $J$ and $K-$band data, our final area totals
$367~\rm{arcmin}^2$. For the $J-$band data, for each field, we
estimated as upper-limit for reliable magnitude measurements the
one corresponding to a $3\sigma$ detection within the circular
aperture adopted by {\tt MAG\_AUTO} for faint and unresolved objects.
Such upper-limit values do not differ significantly from those shown
in Table \ref{maglimJ} as 50\% completeness values.  For each field,
$J-$band magnitudes which were fainter than these values were replaced
by the appropriate upper limits. Figure \ref{JmKvsK} shows the $(J-K)$
\vs $K$ colour-magnitude diagrams for our data.  The objects shown as
star symbols indicate objects classified as point-like using our
star/galaxy classifier (see section \ref{SGsep_sect}).

It is reassuring to see that the majority of objects we classified as
stars are well separated from the global color distribution and are
almost always below the dashed line $(J-K) = 0.96$, corresponding to
the typical color of a main sequence M6 star. Furthermore the stellar
locus is in same position for all four fields, indicating that our
absolute calibration is accurate to within $\sim0.05$ magnitudes. In
Figure \ref{JmKvsK} the dotted line at $(J-K) = 2.3$ corresponds to
colour of an $z \sim 2$ evolved galaxy with a prominent $4000\AA$ break
as a present day elliptical, or a $z\sim0.6$ heavily reddened starburst
galaxy.

\begin{figure}
\centering
\includegraphics[width=8.5cm]{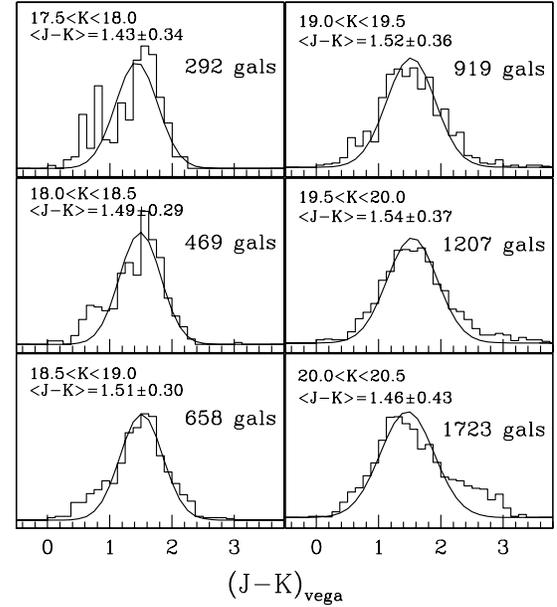} 
\caption{$(J-K)$ colour distribution for the total $K-$selected sample in
  different slices of $K-$band magnitude. Th<ere is a trend towards redder
  colors at fainter magnitudes, probably reflecting an increasing
  fraction of high redshift, red, galaxies.}
\label{colgalJK} 
\end{figure}

Figure \ref{colgalJK} shows galaxy $(J-K)$ color distributions in
different bins of $K-$band magnitudes. It is evident how the population
objects red in $(J-K)$ becomes progressively important at fainter
magnitudes. A detailed analysis of the number counts and clustering
properties of such red population will be presented in a subsequent
paper.

\subsection{Clustering analysis for $K-$selected data}

In this Section we investigate the clustering properties of point like
and extended sources in our $K-$band catalogues.

We use the projected two-point angular correlation function,
$\omega(\theta)$, which measures the excess of pairs separated by an
angle $\theta, \theta+\delta\theta$ with respect to a random
distribution. This statistic is useful for our purposes because it is
particularly sensitive to any residual variations of the magnitude
zero-point across our stacked images. We measure $\omega(\theta)$ using
the standard \citet{Landy_and_Szalay.1993.ApJ} estimator, i.e.,

\begin{equation}
\omega_{e} ( \theta) ={\mbox{DD} - 2\mbox{DR} + \mbox{RR}\over \mbox{RR}}
\label{1.ls}
\end{equation}

with the $\mbox{DD}$, $\mbox{DR}$ and $\mbox{RR}$ terms referring to
the number of data-data, data-random and random-random pairs between
$\theta$ and $\theta + \delta\theta$.  We use logarithmically spaced
bins, with $\Delta log(\theta) = 0.2$, and the angles are expressed in
degrees, unless stated otherwise.  $\mbox{DR}$ and $\mbox{RR}$ are
obtained by populating the two-dimensional coordinate space
corresponding to the different fields number of random points equal to
the number of data points, a process repeated 1000 times to obtain
stable mean values of these two quantities.

\subsubsection {Clustering of point-like sources} 

We first measure the angular correlation function $\omega(\theta)$ of
the stellar sources.  As stars are unclustered, we expect that, if our
magnitude zero-points and detection thresholds are uniform over our
field, then $\omega(\theta)$ should be zero at all angular scales.

\begin{figure}
\centering
\includegraphics[width=8.5cm]{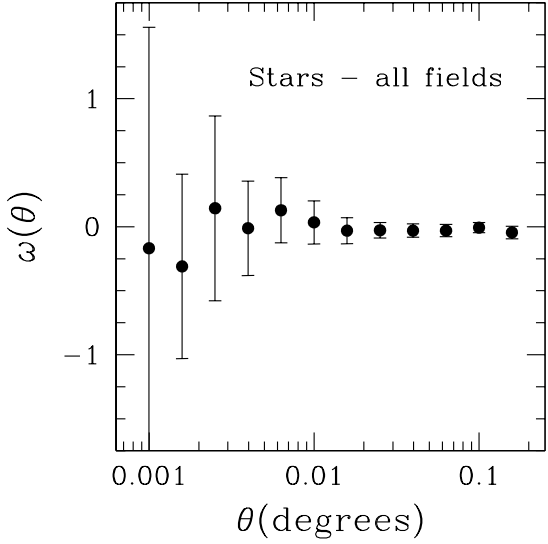} 
\caption{Plot of the correlation function $\omega(\theta)$ for the
total sample of stars in the $K-$band stacks as a function of the
logarithm of the angular pair separation in degrees. At all scales
displayed the measured correlation values are consistent with zero.
}
\label{ksi_stars}
\end{figure}

The results for $K-$band data are displayed in Figure \ref{ksi_stars},
where the correlation function is plotted for the total sample of
stars obtained from all fields according to the procedure described in
section \label{SGsep_sect}: 282 stars in total for the $K-$band
images.  At all scales displayed the measured correlation values are
consistent with zero. Error bars are obtained through bootstrap
resampling of the star sample (and are roughly twice poissonian error
bars).  A similar result is obtained for the measurement of clustering
of stars obtained in $J-$band images.

\subsubsection {Clustering of extended sources}
 
The procedure followed to measure $\omega(\theta)$ is similar to the
one described above for the star sample. In the case of galaxies a
positive amplitude of $\omega(\theta)$ is expected, and we have to take
into account the so called ``integral constraint'' bias.  If the real
$\omega(\theta)$ is assumed to be of the form
$A_{\omega}\theta^{\delta}$, our estimator (\ref{1.ls}) will be offset
negatively from the true $\omega(\theta)$, according to the formula:

\begin{equation}
\omega_e ( \theta) =A_{\omega}(\theta^{-\delta}-C)
\label{2.ls}
\end{equation}

This bias increases as the area of observation decreases, and it is
caused by the need to use the observed sample itself to estimate its
mean density, see {\it eg} \citet{Peebles.1980.LSSU.book}.  The
negative offset AC can be estimated by doubly integrating the assumed
true $\omega(\theta)$ over the field area $\Omega$:

\begin{equation}
A_{\omega}C = {1\over \Omega^2} \int\int \omega ( \theta) d\Omega_1 d\Omega_2
\label{3.ls}
\end{equation}

This integral can be solved numerically using randomly distributed
points for each field:

\begin{equation}
C = {\sum N_{rr}( \theta)~\theta^{-\delta} \over \sum N_{rr}( \theta)}
\label{4.ls}
\end{equation}

Assuming 1'' as the pairs minimal scale at which two galaxies can be
distinguished as separated objects, and $\delta = 0.8$, we obtain the
following values: $C_{0226-04} = 7.58A$, $C_{2217+00} = 9.31A$,
$C_{1003+01} = 8.60A$, $C_{1400+05} = 12.64A$.

We also measured the angular correlation function for our 4 $J-$
selected galaxy samples and verified that they do not exhibit any
significant deviation from a power-law within the angular separation
range associated to our sample areas.

\begin{figure}
\centering
\includegraphics[width=8.5cm]{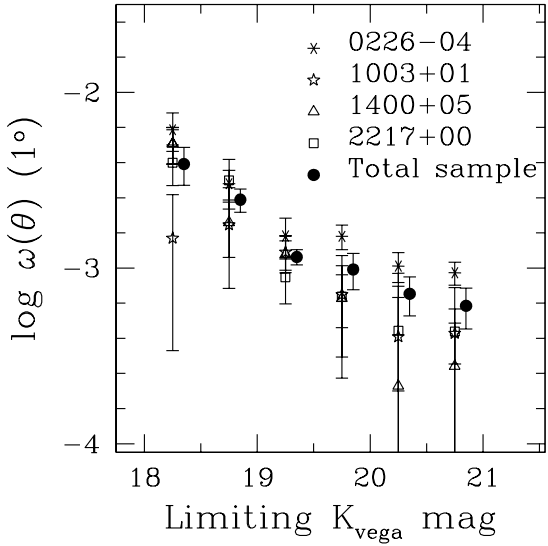} 
\caption{
Results obtained for the amplitude $A_{\omega}$ at $1\deg$ of each of
our fields on galaxy sub-samples of different $K-$band limiting
magnitudes. No correction for stellar contamination is applied and the
error bars on the amplitude are $1\sigma$ error bars obtained from rms
of the fit. The different fields show a good agreement, within the
error bars. On the right of each point, slightly offset for sake of
clarity, filled circles show our final estimate, from the total
sample, of the amplitude of the angular correlation function.
}
\label{ksi_fields}
\end{figure}

We estimated the amplitude $A_{\omega}$ for a series of $K$ limited
galaxy samples by least square fitting $A(\theta^{-0.8}-C)$ to the
observed $\omega(\theta)$, weighting each point using bootstrap error
bars.  Figure \ref{ksi_fields} shows the results obtained for each of
our fields on galaxy sub-samples of different $K-$band limiting magnitudes.
No correction for stellar contamination is applied (only the objects
classified as stars, using the method described in section
\ref{SGsep_sect}, were excluded from the analysis) and the error bars
on the amplitude are $1 \sigma$ error bars obtained from rms of the
fit.  The different fields are in good agreement within the error bars.
The $0226-04$ field is the one with the larger area and the lower
stellar contamination, and this probably explains why its amplitude
values are systematically higher than those of the other fields. It
should be remembered that in the presence of a randomly distributed
spurious population among the sample of objects analized, like faint
stars among our galaxy sample, the resulting measured correlation
amplitudes are reduced by a factor $(1-f)^2$, where $f$ is the fraction
of the randomly distributed component.  Therefore a 10\% (5\%)
contamination rate by stars implies a shifting of the values of the
amplitude plotted by $\sim$ 0.1 (0.06) downwards on the y-axis.
Another point to consider is the expected cosmic variance among
different fields.  A simple numeric estimate of the expected variation
of the measured amplitude of the correlation function on the sky is
$\sigma _A = A_{\omega}^{3/2}C^{1/2}$, see
\citet{Daddi_et_al.2001.A&A}.  For our fields the expected scatter due
to cosmic variance is roughly around 10\% and comparable to the scatter
observed among the fields.

In Figure \ref{ksi_fields} the filled circles show our final estimate
for the amplitude of the correlation function. For each limiting
magnitude these values are obtained by a weighted mean of the
amplitude of our four fields, and their error bars by computing the
weighted variance among amplitudes (weighting proportional to the
error bars on each field value). 

\begin{figure}
\centering
\includegraphics[width=8.5cm]{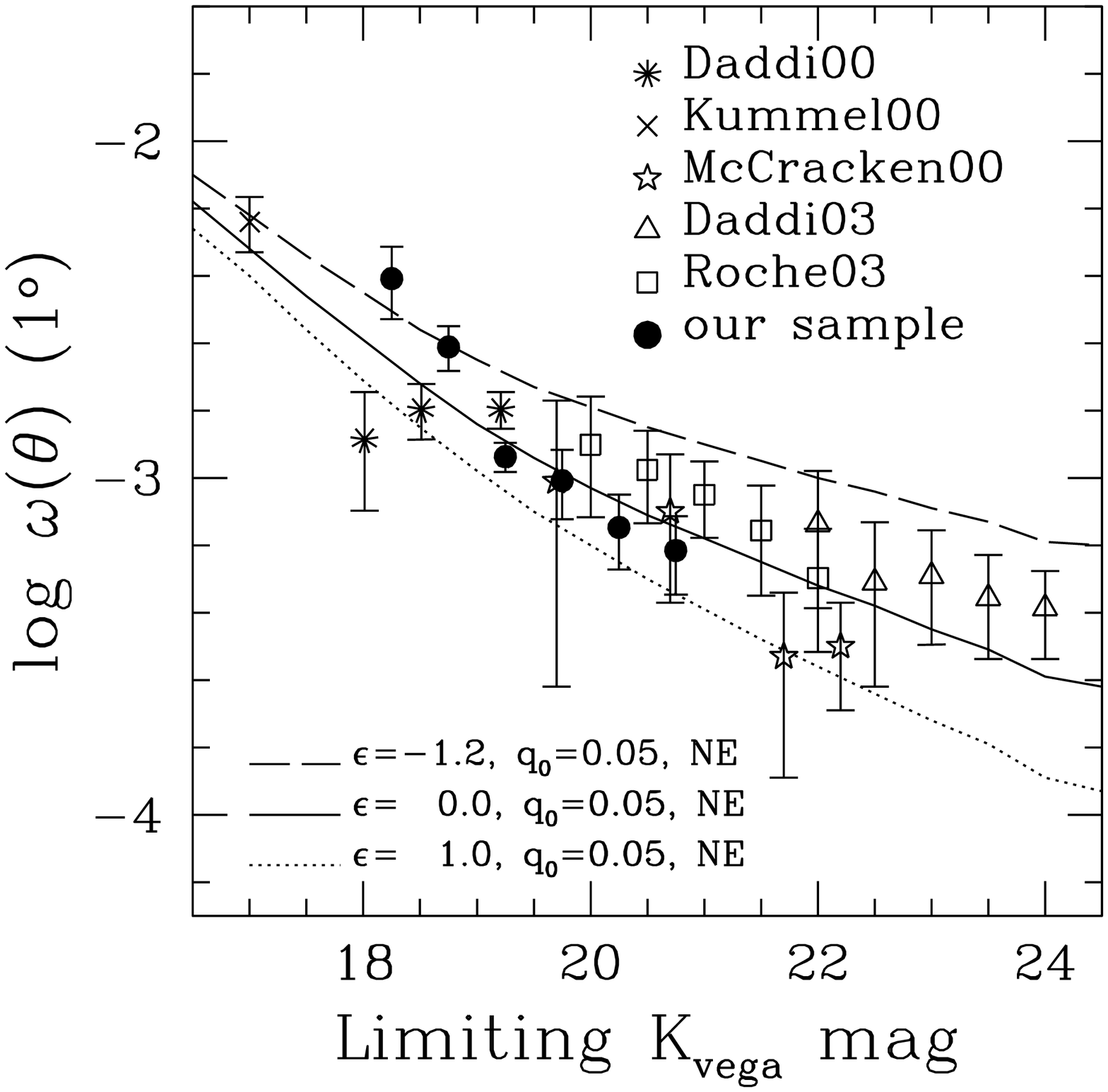} 
\caption{ We compare our total sample estimate of the amplitude,
$A_{\omega}$ at $1\deg$, of the angular correlation function with a
compilation of results from the literature. We include mesurements
from: \citet{Daddi_et_al.2000b.A&A},
\citet{Kummel_and_Wagner.2001A&A},
\citet{McCracken_et_al.2000a.MNRAS}, \citet{Daddi_et_al.2003.ApJ} and
\citet{Roche_et_al.2003.MNRAS}. The continous, dotted and dashed lines
show, for reference, the models of PLE from
\citet{Roche_et_al.1998.MNRAS}, with scaling from local galaxy
clustering. }
\label{ksi_literature}
\end{figure}

Figure \ref{ksi_literature} shows the comparison of our results with
literature data. The continous, dotted and dashed lines show the
models of PLE from \citet{Roche_et_al.1998.MNRAS}, with scaling from
local galaxy clustering.  A more detailed analysis, involving the use
of spectroscopic (from the VVDS redshift survey) and photometric
redshifts for the galaxies of our sample is to be presented in a
forthcoming paper. 

Table \ref{amplitude} lists our amplitude measurements for each field
and for the total sample, in units of $10^{-4}$ at $1\deg$, together with
their $1\sigma$ error bars, computed as explained above.  For each
field and limiting magnitude, the total number of objects N used in
the analysis is also listed.
 
\begin{table*}
\begin{flushleft}
  \begin{tabular}{cccccccccc}
\hline 
Magnitude &\multicolumn {2}{c}{$0226-04$} &\multicolumn {2}{c}{$2217+00$} &\multicolumn {2}{c}{$1003+01$} &\multicolumn {2}{c}{$1400+05$} & Total sample \\
          & N &A $\pm$ dA  &  N     &A $\pm$ dA & N      &A $\pm$ dA& N      &A $\pm$ dA&  A $\pm$ dA      \\ 
\hline 
K $<$ 18.25 & 409 & 61.2  $\pm$ 15.1 & 234  & 39.8 $\pm$ 10.4 & 263 & 14.8 $\pm$ 11.4  & 88  & 51.1  $\pm$ 15.5 & 39.1 $\pm$ 9.5 \\ 
K $<$ 18.75 & 670 & 30.2  $\pm$ 5.8  & 384  & 31.6 $\pm$ 9.9  & 433 & 17.6 $\pm$ 6.1   & 159 & 18.2$\pm$10.5 & 24.5 $\pm$ 3.7 \\
K $<$ 19.25 & 1001 & 15.2 $\pm$ 4.0  & 611  & 8.9  $\pm$ 2.6  & 708 & 12.3 $\pm$ 2.9   & 236 & 12.0$\pm$ 2.3 & 11.6 $\pm$ 1.1 \\
K $<$ 19.75 & 1458 & 15.1 $\pm$ 2.4  & 908  & 6.9  $\pm$ 2.3  & 1056 & 7.1  $\pm$ 4.7  & 343 & 6.7 $\pm$ 3.6 & 9.8  $\pm$ 2.3 \\
K $<$ 20.25 & 2162 & 10.3 $\pm$ 2.0  & 1276 & 4.4  $\pm$ 2.4  & 1506 & 4.1  $\pm$ 3.8  & 488 & 2.1 $\pm$ 2.1 & 7.1  $\pm$ 1.8 \\
K $<$ 20.75 & 3086 & 9.4  $\pm$ 1.4  & 1741 & 4.3  $\pm$ 1.5  & 2149 & 4.2  $\pm$ 3.5  & 709 & 2.8 $\pm$ 2.1 & 6.1  $\pm$ 1.6 \\
            \hline
            \noalign{\smallskip}
          \end{tabular}
        \end{flushleft}
        \caption{Observed $\omega(\theta)$ amplitudes A, in units of
$10^{-4}$ at $1\deg$, together with their $1\sigma$ error bars, for each
of our fields and for the total sample. For each field the number of
objects used in the analysis down to the K magnitude faint limits
shown in the first column is also listed. The number of objects in the
first bin of field $1400+05$ was too small, and therefore this bin was
discarded from the analysis.
}
        \label{amplitude}
\end{table*}

\section{Summary and conclusions}
In this paper we have presented a new near-infrared survey covering
four sub-areas located in each of the four fields of the VIMOS-VLT
deep survey. We have described in detail our data reduction process
starting from pre-reductions, astrometric and photometric
calibrations, image resampling and stacking, and finally extraction of
catalogues. At each stage, we have tried to quantify all sources of
systematic and random errors in our survey. From extensive
simulations, we have shown that our catalogues are reliable in all
fields to at least $K\sim20.75$ and $J\sim21.50$: we define this limit
as the magnitude where object contamination is less than 10\% and
completeness greater than 90\%.

Based on repeated measurements of standard stars, we estimated that
the error on our absolute photometric calibration, field-to-field, is
$\sim 0.05~$magnitudes r.m.s. This is consistent with the $\sim 5\%$
field-to-field variation of $J-$ and $K-$selected galaxy number counts
and the measured field-to-field variation of the $(J-K)$ colour of the
stellar locus. We separated stars from galaxies using the parameter
$r_{1/2}$, which measures the radius for each object which encloses
50\% of the total flux. Stellar counts for our four fields are
consistent with the \citet{Robin_et_al.2003.A&A} model of Milky Way,
and our mean galaxy counts and counts slope over our four fields is
also in excellent agreement with literature compilations. We observe a
change in slope in the $K-$band galaxy number counts at $\sim 17.5$
magnitudes.

We investigated the colour-magnitude distribution of stars and galaxies
identified in our catalogues. All objects lying in the $K$ vs $(J-K)$
stellar locus were successfully identified by our 
classifier. For the galaxy population in the range $17.5<K<20.5$ we
measure a median $(J-K)$ colour of $1.4\pm0.3$, consistent with
published values. This value is remains approximately constant to
progressively fainter magnitudes, until the faintest reliable limits of
our sample ($20.0<K<20.5$). Our fainter magnitude slices show some
evidence of a red tail of objects $(J-K)\sim2.5$ which becomes
progressively larger at fainter magnitudes.

Finally, we measure the angular clustering of stars and galaxies for
our four fields. Our stellar correlation function is consistent with
zero for all four fields on all angular scales. The amplitude of our
galaxy correlation function shows the expected scaling behaviour for
increasingly fainter magnitude slices, and is consistent
previously-presented measurements.

These catalogues will be an excellent tool to investigate the
properties of distant galaxies selected in the near-infrared, and such
investigations which will be the subject of several forthcoming
articles.

\begin{acknowledgements}
This research has been developed within the framework of the VVDS
consortium.\\ This work has been partially supported by the CNRS-INSU
and its Programme National de Cosmologie (France), and by Italian
Ministry (MIUR) grants COFIN2000 (MM02037133) and COFIN2003
(num.2003020150).\\ The VIMOS-VLT observations have been carried out
on guaranteed time (GTO) allocated by the European Southern
Observatory (ESO) to the VIRMOS consortium, under a contractual
agreement between the Centre National de la Recherche Scientifique of
France, heading a consortium of French and Italian institutes, and
ESO, to design, manufacture and test the VIMOS instrument.
H.~J.~McCracken wishes to acknowledge the use of TERAPIX computer
facilities.
\end{acknowledgements}

\bibliographystyle{aa}
\bibliography{2993bib} 

\end{document}